\documentclass[aps,prl,reprint,superscriptaddress]{revtex4-1}
\pdfoutput=1
\usepackage{braket}
\usepackage{color}
\usepackage{graphicx}
\usepackage{verbatim}
\usepackage{bbm}
\usepackage{subcaption}
\usepackage[hidelinks]{hyperref}
\hypersetup{
  colorlinks   = true, 
  urlcolor     = blue, 
  linkcolor    = blue, 
  citecolor   = blue 
}
\usepackage{amsmath}
\usepackage{url}
 \usepackage[normalem]{ulem}

\DeclareMathOperator{\Tr}{Tr}

\begin{document}
\title{Floquet-engineering counterdiabatic protocols in quantum many-body systems}

\author{Pieter~W.\ Claeys}
\email{pwclaeys@bu.edu}
\affiliation{Department of Physics, Boston University, 590 Commonwealth Ave., Boston, MA 02215, USA}

\author{Mohit Pandey}
\affiliation{Department of Physics, Boston University, 590 Commonwealth Ave., Boston, MA 02215, USA}

\author{Dries Sels}
\affiliation{Department of Physics, Harvard University, 17 Oxford St., Cambridge, MA 02138, USA}
\affiliation{Theory of quantum and complex systems, Universiteit Antwerpen, B-2610 Antwerpen, Belgium}

\author{Anatoli Polkovnikov}
\affiliation{Department of Physics, Boston University, 590 Commonwealth Ave., Boston, MA 02215, USA}

\begin{abstract}
Counterdiabatic (CD) driving presents a way of generating adiabatic dynamics at arbitrary pace, where excitations due to non-adiabaticity are exactly compensated by adding an auxiliary driving term to the Hamiltonian. While this CD term is theoretically known and given by the adiabatic gauge potential, obtaining and implementing this potential in many-body systems is a formidable task, requiring knowledge of the spectral properties of the instantaneous Hamiltonians and control of highly nonlocal multibody interactions. We show how an approximate gauge potential can be systematically built up as a series of nested commutators, remaining well-defined in the thermodynamic limit. Furthermore, the resulting CD driving protocols can be realized up to arbitrary order without leaving the available control space using tools from periodically-driven (Floquet) systems. This is illustrated on few- and many-body quantum systems, where the resulting Floquet protocols significantly suppress dissipation and provide a drastic increase in fidelity.
\end{abstract}

\maketitle

\emph{Introduction.} -- Adiabaticity presents one of the fundamental tools in physics, ranging from heat engines in thermodynamics to quantum state preparation and quantum computation \cite{nielsen2000quantum,chandra_quantum_2010,vinjanampathy_quantum_2016,bohn_cold_2017}. However, true adiabatic control can only be obtained using slow driving and asymptotically long time scales. While faster driving leads to diabatic excitations and resulting dissipative losses, the inevitable presence of decoherence and noise in realistic quantum systems limits the available timescales, preventing true adiabaticity. Various methods have been proposed in order to achieve so-called ``Shortcuts to Adiabaticity'' both theoretically \cite{del_campo_shortcuts_2013,torrontegui_chapter_2013,del_campo_focus_2019} and experimentally \cite{bason_highfidelitydriving_2012,zhang_experimentalSTA_2013,an_shortcuts_2016,du_experimentalSTA_2016,zhou_acceleratedsuperadiabatic_2017,wang_experimental_2019,PhysRevLett.122.090502}, mimicking adiabatic dynamics without the requirement of slow driving.

One way of circumventing this loss of fidelity at finite driving rates is through counterdiabatic (CD) or transitionless driving -- a velocity-dependent term is added to the control Hamiltonian, exactly compensating the diabatic contributions to the Hamiltonian in the moving frame \cite{demirplak_adiabatic_2003,demirplak_assisted_2005,berry_transitionless_2009,kolodrubetz_geometry_2017}. This term is known as the \emph{adiabatic gauge potential} (or gauge connection), encoding the geometry of eigenstates when varying a control parameter \cite{kolodrubetz_geometry_2017}. However, while this potential may be exactly obtained in few-body systems, its construction in general requires diagonalization of the Hamiltonian in the full Hilbert space, prohibiting its use in general many-body systems. Furthermore, the resulting operator tends to involve highly nontrivial and nonlocal couplings not present in the control Hamiltonian, preventing its actual implementation (except in some limiting cases) \cite{deffner_classical_2014,zwick_optimized_2014,hatomura_shortcuts_2018,del_campo_assisted_2012,saberi_adiabatic_2014,
del_campo_controlling_2015,okuyama_classical_2016,diao_shortcuts_2018}.

In few-body systems, restricting driving to couplings within the control Hamiltonian led to the development of fast-forward (FF) protocols, where CD driving is effectively realized in a time-dependent rotating frame  \cite{masuda2009fast, torrontegui_shortcuts_2012,bukov_geometric_2018}. However, there exists no general way of constructing these protocols for complex systems. One specific class of FF protocols is those where CD driving is realized through Floquet-engineering: high-frequency oscillations are added to the control so that the resulting Floquet Hamiltonian mimics the CD Hamiltonian. In few-body systems, this has already been used for high-fidelity quantum state manipulation both theoretically in closed \cite{ribeiro_systematic_2017,petiziol_fast_2018,petiziol_accelerating_2019} and open systems \cite{villazon_swift_2019}, and experimentally in a noisy qubit \cite{boyers_floquet-engineered_2018}. 

In this work, we propose a method of (i) finding an efficient and controlled approximation to the gauge potential, remaining well-defined in many-body systems, which can then (ii) be systematically realized through Floquet-engineering by resonantly oscillating the instantaneous Hamiltonian with the driving term. Effectively, we propose a general strategy for designing fast adiabatic protocols, applicable both in small quantum systems to achieve high fidelity for state preparation and in large systems, quantum or classical, to suppress dissipative losses. This is then illustrated on few- and many-body systems. 

\emph{Methods.} -- Consider a control Hamiltonian $\mathcal{H}(\lambda)$ dependent on a single control parameter $\lambda$. Our goal is to transport a stationary state or distribution, at an initial value of the control parameter $\lambda_i$, to one corresponding to a final value $\lambda_f$.  In the standard approach, this is done by adiabatically changing $\lambda(t)$ from $\lambda_i$ to $\lambda_f$, which is often impractical because of the necessary access to long timescales. The key idea of CD driving is to vary the parameter $\lambda(t)$ at a finite rate while simultaneously compensating the diabatic excitations by explicitly adding an auxiliary term as
\begin{equation}\label{eq:CD_driving}
\mathcal{H}_{CD}(t) = \mathcal{H}(\lambda)+ \dot{\lambda} \mathcal{A}_{\lambda}.
\end{equation}
Adiabatic control at arbitrary driving rates and for arbitrary initial states is realized provided the adiabatic gauge potential $\mathcal{A}_{\lambda}$ \cite{kolodrubetz_geometry_2017} satisfies
\begin{equation}\label{eq:AGP:DefMatEl}
\braket{m|\mathcal{A}_{\lambda}|n} = i \braket{m|\partial_{\lambda}n} = -i \frac{\braket{m|\partial_{\lambda} \mathcal{H}|n} }{\epsilon_m-\epsilon_n},
\end{equation}
where $\ket{n}$ and $\epsilon_n$ are the eigenstates and the energy spectrum of the instantaneous Hamiltonian, $\mathcal{H}(\lambda)\ket{n} = \epsilon_n \ket{n}$. The  CD term then exactly compensates non-adiabatic transitions between eigenstates.

The expression~\eqref{eq:AGP:DefMatEl} already highlights the issues with many-body CD driving: since the gauge potential is defined in the eigenbasis of the instantaneous Hamiltonian, it requires exact diagonalization. Furthermore, for increasing system sizes the denominator $(\epsilon_m-\epsilon_n)$ can become exponentially small, leading to divergent matrix elements and an ill-defined gauge potential in the thermodynamic limit \cite{jarzynski_geometric_1995,kolodrubetz_geometry_2017}. Physically, at least in chaotic systems, the exact gauge potential also cannot  be local because no local operator is expected to be able to distinguish general many-body states with arbitrary small energy difference \cite{dalessio_quantum_2016}. Considering a system with a gapped ground state, Lieb-Robinson bounds can however be used to obtain a quasi-local operator reproducing the action of the exact gauge potential on this ground state, since no such divergences occur in this case \cite{bachmann_adiabatic_2017}.

In the following, we propose a general approximate gauge potential defined as
\begin{equation}\label{eq:com:def_expansion}
\mathcal{A}^{(\ell)}_{\lambda} = i \sum_{k=1}^{\ell} \alpha_k \underbrace{[\mathcal{H},[\mathcal{H},\dots [\mathcal{H}}_{2k-1}, \partial_{\lambda}\mathcal{H}]]],
\end{equation}
fully determined by a set of coefficients $\{\alpha_1, \alpha_2, \dots, \alpha_{\ell}\}$, where $\ell$ determines the  order of the expansion. It can be shown  that the exact gauge potential can be represented in this form in the limit $\ell\to\infty$ \cite{Note1}. Instead we consider a small finite value of $\ell$ and treat the expansion coefficients as variational parameters, which can be obtained by minimizing the action $S_{\ell}$
\begin{equation}
S_{\ell} = \Tr \left[G_{\ell}^2\right], \qquad  G_{\ell} = \partial_{\lambda}\mathcal{H}-i[\mathcal{H},\mathcal{A}_{\lambda}^{(\ell)}].
\end{equation}
The exact gauge potential is known to follow from the variational minimization of an action \cite{sels_minimizing_2017}. However, it is not a priori clear what (local) operators should be included in the variational basis. The total number of possible operators increases exponentially with their support, limiting the brute-force minimization to highly local operators with restricted support. Furthermore, it is far from guaranteed that such operators will be experimentally realizable. The main finding of the present work is that the proposed ansatz tackles both problems simultaneously. (i) The number of variational coefficients can be kept small while still returning an accurate approximation to the exact gauge potential. As such, Eq.~(\ref{eq:com:def_expansion}) can be seen as a variational ansatz including only the most important contributions with the maximum range of operators set by $\ell$. (ii) In addition, this gauge potential can be engineered with a simple Floquet protocol. Essentially, this realization is possible because the high-frequency expansion of the Floquet Hamiltonian shares the commutator structure of Eq.~\eqref{eq:com:def_expansion}. This expansion exhibits the symmetries of the exact solution at each order, and as additional bonus we point out that this ansatz has a well-defined classical limit, where even the local-operator basis becomes infinite-dimensional. In classical systems, the commutators in Eq.~\eqref{eq:com:def_expansion} only need to be replaced by Poisson brackets.

Since the action is simply the Hilbert-Schmidt norm of $G_{\ell}$, the variational method has the clear advantage that the action can be calculated without explicitly constructing the operator matrix in the full Hilbert space. There are various ways of motivating Eq.~(\ref{eq:com:def_expansion}) (see Supplementary Material \footnote{Supplementary Material.} for more details): it can be seen as an expansion in the Krylov subspace generated by the action of $G_{\ell}$, or by noting that such commutators appear through the Baker-Campbell-Hausdorff expansion in the definition of a (properly regularized) gauge potential, or by simply noting that its matrix elements share the general structure of those of the exact gauge potential. Namely, evaluating Eq.~(\ref{eq:com:def_expansion}) in the eigenbasis of $\mathcal{H}$ returns
\begin{align}\label{eq:MatEl}
\braket{m|\mathcal{A}^{(\ell)}_{\lambda}|n} &= i \sum_{k=1}^{\ell} \alpha_k \braket{m| \underbrace{[\mathcal{H},[\mathcal{H},\dots [\mathcal{H}}_{2k-1}, \partial_{\lambda}\mathcal{H}]]]|n} \nonumber \\
&= i \left[\sum_{k=1}^{\ell}\alpha_k (\epsilon_m-\epsilon_n)^{2k-1}\right] \braket{m|  \partial_{\lambda}\mathcal{H} |n}.
\end{align}
This can be compared to the exact expression (\ref{eq:AGP:DefMatEl}), containing a state-dependent factor $\braket{m|\partial_{\lambda}\mathcal{H}|n}$ and a prefactor only dependent on the excitation frequency $\omega_{mn} = (\epsilon_m-\epsilon_n)$. The variational optimization can then be seen as approximating the exact prefactor $1/\omega_{mn}$ by a power-series prefactor $a_{\lambda}^{(\ell)}(\omega_{mn}) \equiv \sum_{k=1}^{\ell} \alpha_k \omega_{mn}^{2k-1}$ for the range of relevant excitation frequencies set by $\braket{m|  \partial_{\lambda}\mathcal{H} |n}$.

While such an approximation is generally impossible due to the divergence of $1/\omega_{mn}$ near $\omega_{mn}=0$ and the divergence of the power series for $\omega_{mn} \to \infty$, the approximation does not need to hold in these limits. First, for large $\omega_{mn}$ the matrix elements of local operators $\braket{m|\partial_{\lambda}\mathcal{H}|n}$ typically decay exponentially with $\omega_{mn}$~\cite{dalessio_quantum_2016}, leading to a negligible contribution to the gauge potential. Second, there are physical motivations for allowing transitions for small $\omega_{mn}$. When speeding up adiabatic driving in the presence of an energy gap $\Delta$, only transitions with $\omega_{mn} \geq \Delta$ need to be suppressed in order to achieve unit fidelity, and in more general gapless regimes corresponding to e.g. excited states the resulting excitations will be confined to a narrow energy shell, the width of which decreases with the order $\ell$ of the expansion.

\begin{figure}[ht!]
\centering
\includegraphics[width=\columnwidth]{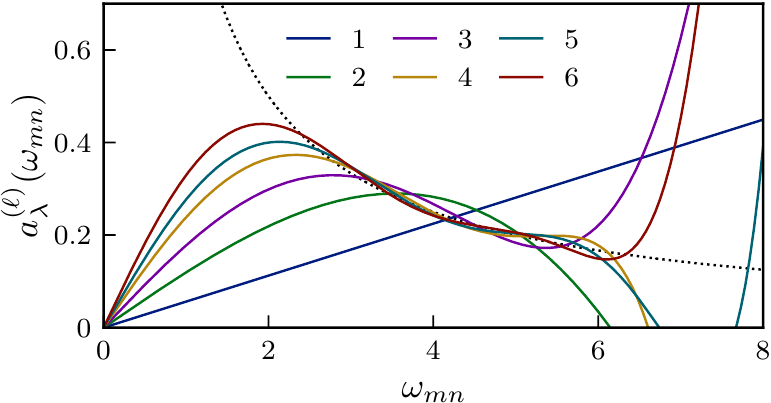}
\caption{Variationally-obtained power-series prefactor $a_{\lambda}^{(\ell)}(\omega_{mn})$ for Eq.~(\ref{eq:Ising}). Dotted line corresponds to exact prefactor $1/\omega_{mn}$. Parameters $L=14$, $J=1$, $h_x = h_z = 0.3$, $\lambda=1$.\label{fig:fitting}}
\end{figure}

We illustrate how this expansion works in Fig.~\ref{fig:fitting}, for a non-integrable Ising chain with
\begin{align}\label{eq:Ising}
\mathcal H= J \sum_{i=1}^L \sigma_i^z \sigma_{i+1}^z  + \lambda \left( h_z \sum_{i=1}^L \sigma_i^z +  h_x \sum_{i=1}^L \sigma_i^x\right),
\end{align}
where no exact gauge potential can be obtained in the thermodynamic limit. It is clear that the variational optimization returns a gauge potential optimized for a relevant window of excitation frequencies, where the approximation necessarily improves with increasing $\ell$.

The resulting gauge potential can immediately be used to reliably speed up adiabatic protocols by considering a driving protocol $\mathcal{H}^{(\ell)}_{CD}(t) = \mathcal{H}(\lambda) + \dot{\lambda} \mathcal{A}_{\lambda}^{(\ell)}(\lambda)$. While this presents a guaranteed improvement in fidelity, it also requires access to interaction terms not necessarily available within the protocol, where the only interactions that are generally present are those of $\mathcal{H}(\lambda)$ and $\partial_{\lambda}\mathcal{H}(\lambda)$. Remarkably, this CD Hamiltonian can also be realized as an effective Floquet Hamiltonian by simply oscillating these two terms at high frequency. Consider
\begin{align}\label{eq:H_FE}
\mathcal{H}_{FE}(t) =& \left[1+\frac{\omega}{\omega_0} \cos(\omega t) \right]\mathcal{H}(\lambda) \nonumber \\
&+\dot{\lambda}\left[\sum_{k=1}^{\infty}\beta_{k} \sin\left((2k-1)\omega  t\right)\right] \partial_{\lambda}\mathcal{H}(\lambda),
\end{align}
with $\beta_k$ the Fourier coefficients of the additional drive and $\omega_0$ a reference frequency, both of which will be determined later. Floquet theory then allows for the definition of a time-independent Floquet Hamiltonian reproducing time evolution over a single driving cycle (with $T = 2\pi/\omega$)
\begin{align}
 \exp \left( -i \mathcal{H}_F  T \right) \equiv \mathcal{T} \exp \left(- i \int_{t}^{t+T} \mathcal{H}_{FE}(t')\ dt' \right).
\end{align}
The limit where the driving term scales with the frequency is known to give rise to non-trivial Floquet Hamiltonians $\mathcal{H}_F$ in various scenarios \cite{goldman_periodically_2014,goldman_periodically_2015,bukov_universal_2015}, perhaps most importantly in dynamical decoupling \cite{mentink_manipulating_2017,claassen_dynamical_2017}. In the same way that Floquet driving can be used to reduce interactions within a Hamiltonian, this can also be used to reduce excitations within the current protocol. 

More specifically, the proposed series expansion for the adiabatic gauge potential can be implemented in the infinite-frequency limit $\omega \to \infty$, realizing (stroboscopic) CD driving. This Floquet Hamiltonian can be obtained from the Magnus expansion, presenting a series expansion of $\mathcal{H}_F$ in powers of the inverse-frequency. Essentially, the $\omega \to \infty$ limit combined with the scaling of $\mathcal{H}$ with $\omega$ guarantees that only commutators of the form $[\mathcal{H}, \dots, [\mathcal{H},\partial_{\lambda}\mathcal{H}]]]$ survive in the Magnus expansion, which can then be found as $\mathcal{H}_F = \mathcal{H}(\lambda)+\dot{\lambda}\mathcal{A}_F$ \cite{Note1}, with
\begin{equation}
\braket{m|{\mathcal{A}}_F|n} = i\sum_{k=1}^{\infty}\beta_k \mathcal{J}_{2k-1} \left(\frac{\omega_{mn}}{\omega_0} \right) \braket{m|\partial_{\lambda}\mathcal{H}|n},
\end{equation}
where $\mathcal{J}_{k}$ are Bessel functions of the first kind. Again, this reproduces the correct structure of the gauge potential, where the frequency-dependent prefactor is now expressed in terms of $\mathcal{J}_{k}$. For small $\omega_{mn}/\omega_0$, $\mathcal{J}_k(\omega_{mn}/\omega_0) \propto \omega_{mn}^{k}$, which can be used to stroboscopically engineer the CD term by choosing the Fourier harmonics in such a way that the Floquet prefactor reproduces the power series (\ref{eq:MatEl}) in the relevant range of excitation frequencies. In first approximation, this can be done by restricting time-evolution to $\ell$ harmonics and setting
\begin{align}
\sum_{k=1}^{\ell} \beta_k  \mathcal{J}_{2k-1}\left(\frac{\omega_{mn}}{\omega_0}\right)  &= \sum_{k=1}^{\ell} \alpha_k \omega_{mn}^{2k-1}  + \mathcal{O}(\omega_0^{-2}).
\end{align}
Analytic expressions can easily be obtained for matching the harmonics to the coefficients in the gauge potential up to arbitrary order and, if necessary, higher-order harmonics can be added in order to compensate the $\mathcal{O}(\omega_0^{-2})$ corrections order by order \cite{Note1}. 

As an illustration, considering the expansion for a series with two harmonics leads to
\begin{align}
\mathcal{H}_F &= \mathcal{H}({\lambda})+\frac{i\dot{\lambda}}{2}\frac{\beta_1}{\omega_0}  [\mathcal{H},\partial_{\lambda}\mathcal{H}]  \nonumber\\
&+\frac{i\dot{\lambda}}{48}\frac{\left(\beta_2-3\beta_1\right)}{\omega_0^3} [\mathcal{H},[\mathcal{H}, [\mathcal{H},\partial_{\lambda}\mathcal{H}]]]+\mathcal{O}(\omega_0^{-5}).
\end{align}
Then choosing $\beta_1 = 2 \alpha_1 \omega_0$ and $\beta_2 = 2\omega_0(24 \alpha_2\omega_0^2+3 \alpha_1)$, the Floquet Hamiltonian following from the driving (\ref{eq:H_FE}) can be matched to the Eqs. (\ref{eq:CD_driving}) and (\ref{eq:com:def_expansion}), returning
\begin{align}
\mathcal{H}_F =&\mathcal{H}+ i \dot{\lambda} \alpha_1 [\mathcal{H},\partial_{\lambda}\mathcal{H}] \nonumber \\
&+ i \dot{\lambda} \alpha_2 [\mathcal{H},[\mathcal{H}, [\mathcal{H},\partial_{\lambda}\mathcal{H}]]]+\mathcal{O}(\omega_0^{-2}).
\end{align}
This protocol approximately reproduces the CD evolution at stroboscopic times $t =n \cdot T, n \in \mathbbm{N}$. Note that, while this protocol does not introduce new interactions in the Hamiltonian, the additional cost is that it requires oscillations of both $\mathcal{H}$ and $\partial_{\lambda}\mathcal{H}$ rather than just $\partial_{\lambda}\mathcal{H}$.

\begin{figure}[ht]
\includegraphics[width=\columnwidth]{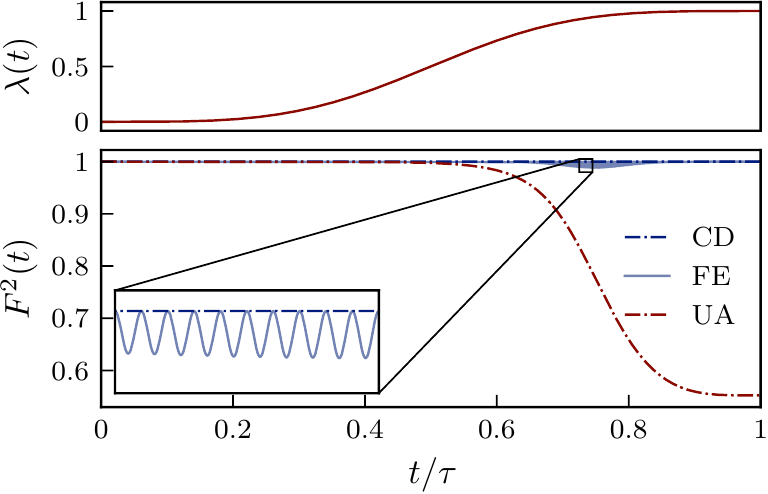}%
\caption{Fidelity in the 2-qubit system (\ref{eq:2LS}) for UA, CD and FE protocol. Increasing $\omega$ further suppresses the Floquet oscillations. Parameters $J=-1$, $h_z=5$, $\tau = 0.1$, $\omega_0 =  10 \cdot 2\pi $ and $\omega = 250 \cdot  \omega_0$. \label{fig:2qubit}}
\vspace{-\baselineskip}
\end{figure}

\emph{Applications.} -- This procedure can now be applied on various systems with increasing complexity. In all examples, we will consider a specific driving protocol with
\begin{equation}
\lambda(t) = \sin^2\left(\frac{\pi}{2}\sin^2\left(\frac{\pi t}{2 \tau}\right)\right),
\end{equation}
ramping from $\lambda(0)=0$ to $\lambda(\tau)=1$ in such a way that $\dot{\lambda}$ and $\ddot{\lambda}$ vanish at the beginning and end of the protocol. $\lambda$ then behaves as an annealing parameter, and as first measure for the effectiveness of the protocol we will initialize the system in the ground state for $\lambda=0$ and calculate the fidelity of the time-evolved state w.r.t. the instantaneous ground state $F^2(t)=|\braket{\psi(t)|\psi_0(\lambda(t))}|^2$.

First consider a two-qubit system, for which all calculations can be performed analytically \cite{Note1},
\begin{equation}\label{eq:2LS}
\mathcal{H}(\lambda) = J \left(\sigma_1^x\sigma_2^x+\sigma_1^z\sigma_2^z\right) + h_z (\lambda-1)\left(\sigma_1^z+\sigma_2^z\right).
\end{equation}
The first-order expansion leads to
\begin{equation}
\mathcal{A}_{\lambda}^{(1)} = -\frac{ J h_z}{2}\frac{\left(\sigma_1^y\sigma_2^x+\sigma_1^x\sigma_2^y\right)}{J^2+4(\lambda-1)^2h_z^2}  .
\end{equation}
Remarkably, this  already returns the exact adiabatic gauge potential as presented in Ref.~\cite{petiziol_fast_2018}. This can be understood either by noting that $[\mathcal{H},[\mathcal{H},[\mathcal{H}, \partial_{\lambda} \mathcal{H} ]] \propto  [\mathcal{H},\partial_{\lambda} \mathcal{H}]$, such that the higher-order commutators do not introduce new operators in the expansion, $\mathcal{A}_{\lambda}^{(\ell)} \propto \mathcal{A}_{\lambda}^{(1)}$, and the variational approach can be seen as a resummation of all higher-order terms exactly determining the prefactor. Second, this system behaves as a two-level system since any instantaneous Hamiltonian only couples $\ket{\downarrow \downarrow}$ and $\ket{\uparrow \uparrow}$, leading to a single excitation frequency which can be exactly cancelled by a single commutator.

\begin{figure}[ht]
\includegraphics[width=\columnwidth]{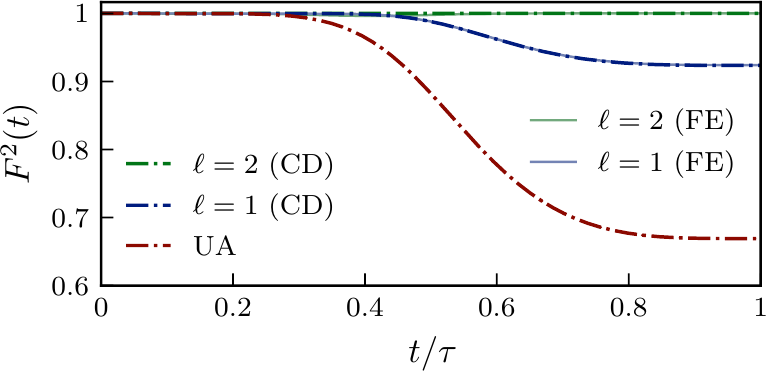}%
\caption{Fidelity in the 2-qubit system (\ref{eq:3LS}) for the UA, CD and FE protocol with $\ell=1,2$. Parameters $\tau = 0.1$, $J=1$, $h=2$, $\omega_0 =  10 \cdot 2\pi $ and $\omega = 2.5 \cdot 10^{2\ell} \cdot  \omega_0$. \label{fig:3LS}}
\vspace{-\baselineskip}
\end{figure}

The resulting CD driving can be realized up to $\mathcal{O}(\omega_0^{-2})$ using a single harmonic as
\begin{align}
&\mathcal{H}_{FE}(t) = \left[1+\frac{\omega}{\omega_0} \cos(\omega t) \right] \mathcal{H}(\lambda(t)) \nonumber\\ 
&\qquad -  \dot{\lambda} \frac{2 h_z \omega_0 \sin(\omega t)}{4J^2+16(\lambda(t)-1)^2h_z^2} \left(\sigma_1^z+\sigma_2^z\right).
\end{align}
The results are illustrated in Fig.~\ref{fig:2qubit}, where the duration of the protocol has been chosen in such a way that $\tau$ is too small for the unassisted (UA) protocol to accurately prepare the final Bell state $\ket{\psi_0(\lambda=1)} = \frac{1}{\sqrt{2}}(\ket{\uparrow \uparrow}+\ket{\downarrow \downarrow})$. Exact CD driving returns unit fidelity by definition, which can be well approximated (with a final error of the order $10^{-5}$) using the proposed Floquet-engineered (FE) protocol. 

Next, consider a two-qubit system behaving as a three-level system,
\begin{equation}\label{eq:3LS}
\mathcal{H}(\lambda) = -2J \sigma_1^z \sigma_2^z - h\left(\sigma_1^z+\sigma_2^z\right)+2 h \lambda \left(\sigma_1^x+\sigma_2^x\right),
\end{equation}
where the total spin-0 state $\ket{\uparrow \downarrow}-\ket{\downarrow \uparrow}$ decouples from the rest of the Hilbert space. Transitionless protocols in three-level systems have been a recent subject of interest \cite{martinez-garaot_shortcuts_2014,song_physically_2016}, since the exact gauge potential can no longer be trivially obtained. As shown in Fig.~\ref{fig:3LS}, the fidelity for the unassisted protocol is $67\%$, increasing to $92\%$ for $\ell=1$, before reaching approximate unit fidelity (up to an error $10^{-6}$) for $\ell=2$. Again, for $\ell=2$  the variational approach returns the exact gauge potential, without any reference to exact diagonalization, since only two excitation frequencies are present in the system. The FE protocol accurately reproduces the CD protocol.

\emph{Magnetic trap.} -- Moving to many-body systems, we consider the non-integrable Ising chain. Rather than simply changing the magnetic field uniformly, we will consider a more involved protocol where a local Gaussian magnetic trap is moved across the chain, similar to the `optical tweezers' problem \cite{sorensen_exploring_2016}. In this problem, a set of initially localized spins are to be moved across the model while minimizing dissipation. The full Hamiltonian is given by
\begin{align}
&\mathcal{H}(\lambda) = \mathcal{H}_{0}-h_t \sum_{i=1}^L  \exp\left[{-\frac{(i-c_t(\lambda))^2}{w_t^2}}\right] \sigma_i^z, \\
&\mathcal{H}_0=J \sum_{i=1}^{L-1} \sigma_i^z \sigma_{i+1}^z  + h_z \sum_{i=1}^L \sigma_i^z + h_x \sum_{i=1}^L \sigma_i^x, 
\end{align}
with $c_t(\lambda) = (1-\lambda)i_0 +\lambda i_f$. Tuning $\lambda$ from $0$ to $1$ then drags the center of the trap $c_t(\lambda)$  with strength $h_t$ and width $w_t$ from site $i_0$ to $i_f$.

\begin{figure}[ht]
  \centering
  \begin{subfigure}{\linewidth}
    \centering
    \includegraphics[width=\columnwidth]{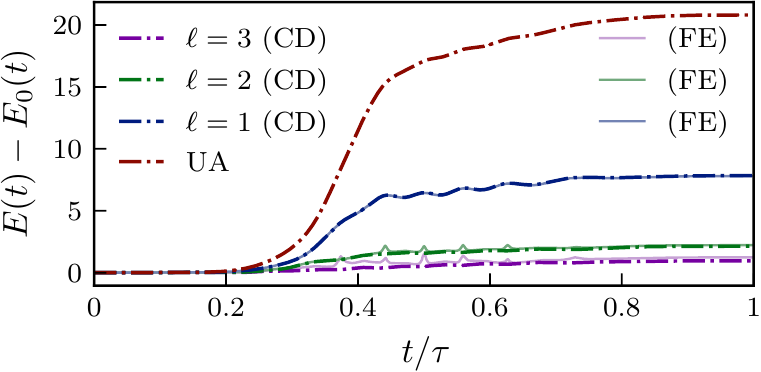}
    \caption{Absorbed energy for the UA, CD and FE protocol with $\ell=1,2,3$.\label{fig:trap:diss}}
  \end{subfigure}

  \begin{subfigure}{\linewidth}
    \centering
    \includegraphics[width=\columnwidth]{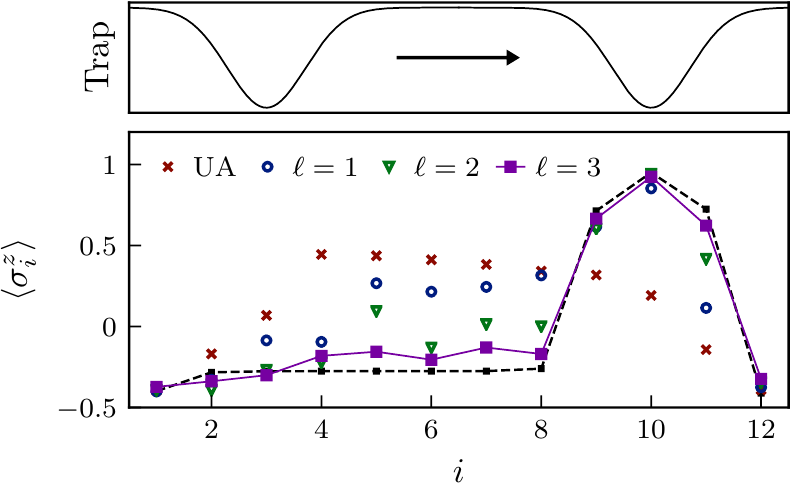}
    \caption{Spin profile at time $\tau$ for the UA and the CD protocol with $\ell=1,2,3$. Exact final profile is given in a dashed line.\label{fig:trap:spin}}
  \end{subfigure}  
  \caption{Moving the magnetic trap in time $\tau = 0.5$ from site $n_0=3$ to site $n_f=10$ for an Ising model with parameters $L=12$, $J=-1$, $h_x = 0.8$, $h_z=0.9$, $h_t=8$, $w_t=1$. $\omega_0 = 10 \cdot 2 \pi$ and $\omega = 10^{4} \cdot \omega_0$. \label{fig:magnTrap}}  
\vspace{-2\baselineskip}
\end{figure}  

Rather than the fidelity, we now consider the absorbed energy $E(t)-E_0(t)=\braket{\psi(t)|\mathcal{H}(\lambda(t))|\psi(t)}-\braket{\psi_0(t)|\mathcal{H}(\lambda(t))|\psi_0(t)}$ as a measure for the dissipation in the system, as shown in Fig.~\ref{fig:trap:diss} for $\ell=1,2,3$. It is clear that, for the given protocol duration, the UA protocol fails in reproducing the final state. This is then remedied by including the CD terms with $\ell=1,2,3$, reducing the dissipation and absorbed energy by a factor 20 \footnote{This corresponds to an increase in the final fidelity from $2.8\%$ to $90\%$}. The Floquet-engineered drive succeeds in reproducing the CD results, with only minor deviations at intermediate times when $E_0(t)$ becomes extremal. The improved performance can also be observed in the final spin profile of $\sigma_i^z$ (Fig.~\ref{fig:trap:spin}), where the CD driving is crucial in reproducing the exact result. While the proposed method seems to work particularly well for this type of model, as also observed in the optical case \cite{sels_stochastic_2018}, this is representative for more general many-body systems.

Finally, note that in this calculation it was not the derivation of the gauge potential and the Floquet drive that was the bottleneck, but rather the time evolution as validation of the protocol. The former remain applicable for arbitrary large system sizes and should similarly lead to significant suppression of energy losses.

\emph{Conclusion and outlook}. -- In this work, it was argued that the adiabatic gauge potential can be efficiently constructed as a series of variationally-optimized nested commutators. This expansion can be constructed without having to resort to exact diagonalization and remains well-defined in many-body systems. Due to the similarity between this series and the Magnus expansion in periodically-driven systems, this expansion is easily realized through Floquet-engineering, such that the resulting approximate counterdiabatic/transitionless driving protocol can be realized via Floquet driving without introducing additional terms in the Hamiltonian. As illustrated on two-qubit systems and a non-integrable Ising chain, a small number of terms can already result in a drastic increase in fidelity in few- and many-body systems. This presents the usual trade-off in fast-forward protocols, where an increase in fidelity can be obtained provided precise control over the driving and access to large interaction strengths is available \cite{demirplak_consistency_2008,funo_universal_2017,zheng_cost_2016}.

Future applications and extensions are plenty. First, while all current simulations were performed on spin systems, the method can immediately be extended towards bosonic or fermionic models. Second, while the presented expansion of the gauge potential is particularly convenient for CD driving (where only a single state is involved), the exact gauge potential contains extensive information about the geometry of all quantum states, adiabatic deformations, integrability and its violations, approximate conservation laws and many other properties, which can also be extracted from the current approximation. These methods should also allow for the construction of approximately-conserved operators, and the similarity of the proposed expansion to the Magnus expansion allows for the realization of integrable gauge potentials in analogy with integrable Floquet Hamiltonians \cite{gritsev_integrable_2017}.

\section*{Acknowledgments}
P.W.C. gratefully acknowledges a Francqui Foundation Fellowship from the Belgian American Educational Foundation and support from Boston University's Condensed Matter Theory Visitors program. M.P acknowledges support from Banco Santander Boston University-National University of Singapore grant. D.S. acknowledges support from the FWO as postdoctoral fellow of the Research Foundation-Flanders. A.P was supported by NSF DMR-1813499 and AFOSR FA9550-16-1-0334. Calculations were performed using QuSpin \cite{weinberg_quspin:_2017,weinberg_quspin:_2018}, and we acknowledge Jonathan Wurtz for providing code for the variational optimization of adiabatic gauge potentials. This work benefited from discussions with Anatoly Dymarsky and Tamiro Villazon.

\bibliographystyle{apsrev4-1}
\bibliography{MyLibrary.bib}
\end{document}


\title{Supplementary Material}

\author{Pieter~W.\ Claeys}
\email{pwclaeys@bu.edu}
\affiliation{Department of Physics, Boston University, 590 Commonwealth Ave., Boston, MA 02215, USA}

\author{Mohit Pandey}
\affiliation{Department of Physics, Boston University, 590 Commonwealth Ave., Boston, MA 02215, USA}

\author{Dries Sels}
\affiliation{Department of Physics, Harvard University, 17 Oxford St., Cambridge, MA 02138, USA}
\affiliation{Theory of quantum and complex systems, Universiteit Antwerpen, B-2610 Antwerpen, Belgium}

\author{Anatoli Polkovnikov}
\affiliation{Department of Physics, Boston University, 590 Commonwealth Ave., Boston, MA 02215, USA}

\maketitle

\section{Regularized gauge potentials}
As one way of motivating the variational ansatz, the adiabatic gauge potential (for fixed $\lambda$) can be rewritten as
\begin{align}
\mathcal{A}_{\lambda} &= \lim_{\epsilon \to 0^+} \int_{0}^{\infty} dt  \ e^{-\epsilon t} \left( e^{-i \mathcal{H}(\lambda) t} \partial_{\lambda}\mathcal{H}(\lambda) e^{i\mathcal{H}(\lambda)t} -\mathcal{M}_{\lambda}\right) 
\end{align}
with $\mathcal{M}_{\lambda}=\overline{\partial_{\lambda }\mathcal{H}} = \sum_n \ket{n}\braket{n|\partial_{\lambda} \mathcal{H} |n}\bra{n}$ cancelling the diagonal elements by construction, similar to the integral expression for the classical gauge potential \cite{jarzynski_geometric_1995}. This immediately follows from the evaluation of the off-diagonal elements
\begin{align}
\braket{m|\mathcal{A}_{\lambda}|n} &= \lim_{\epsilon \to 0^+} \int_{0}^{\infty} dt \  e^{-\epsilon t}  e^{-i (\epsilon_m-\epsilon_n) t}  \braket{m|\partial_{\lambda}\mathcal{H}|n} \nonumber \\
&= \lim_{\epsilon \to 0^+} \frac{\braket{m|\partial_{\lambda}\mathcal{H}|n}}{\epsilon+i (\epsilon_m-\epsilon_n)}.
\end{align}
From the Baker-Campbell-Hausdorff expansion, we can write
\begin{equation}
 e^{-i \mathcal{H} t} \partial_{\lambda}\mathcal{H} e^{i\mathcal{H}t} = \sum_{k=0}^{\infty} \frac{(-it)^k}{k!}\underbrace{[\mathcal{H},[\mathcal{H},\dots [\mathcal{H}}_{k}, \partial_{\lambda}\mathcal{H}]],
\end{equation}
where the (real for real Hamiltonians) even-order commutators will contribute to $\mathcal{M}_{\lambda}$ and the odd-order commutators constitute $\mathcal{A}_{\lambda}$. While the resulting geometric series is not convergent for small $\epsilon$, this hints at the use of the nested commutators to reconstruct the gauge potential.

A `gapped' gauge potential $\mathcal{A}_{\lambda}^{\Delta}$ can alternatively be defined as
\begin{equation}
\mathcal{A}_{\lambda}^{\Delta} \equiv  i \sum_{k=1}^{\infty} \frac{(-1)^k}{k!} \Delta^{-2k}\underbrace{[\mathcal{H},[\mathcal{H},\dots [\mathcal{H}}_{2k-1}, \partial_{\lambda}\mathcal{H}]],
\end{equation}
satisfying
\begin{align}
\braket{m|\mathcal{A}_{\lambda}^{\Delta}|n} &= i \sum_{k=1}^{\infty} \frac{(-1)^k}{k!} \Delta^{-2k} (\epsilon_m-\epsilon_n)^{2k-1}\braket{m|\partial_{\lambda} \mathcal{H}|n}\nonumber \\
&= -i\frac{1-e^{-(\epsilon_m-\epsilon_n)^2/\Delta^2}}{\epsilon_m-\epsilon_n} \braket{m|\partial_{\lambda} \mathcal{H}|n},
\end{align}
acting as the exact gauge potential for all excitation frequencies $(\epsilon_m-\epsilon_n)$ above a gap $\Delta$, and vanishing for excitation frequencies $(\epsilon_m-\epsilon_n)$ below $\Delta$. Through the introduction of a finite gap, a regularized gauge potential can be expressed in terms of nested commutators, remaining well-defined in the thermodynamic limit, which can then be used to strongly suppress excitations above this gap, similar in spirit to Ref.~\cite{bachmann_adiabatic_2017}. In practice, this series summation will be truncated, where the variational minimization is guaranteed to return the optimal series coefficients.

Alternatively, the variational optimization can be avoided if we only wish to approximate the prefactor $1/\omega_{mn}$ for a given range of excitation frequencies $\omega_{mn}\in [\Delta_{min},
\Delta_{max}]$ for a given $\Delta_{min}$ and $\Delta_{max}$. This could occur in systems with a known gap or a given excitation spectrum, where the counterdiabatic driving only needs to suppress excitations in a known frequency window. In this case, the fitting implicit in the variational procedure can be replaced by a straightforward fitting of $1/\omega_{mn}$ to a power series of a given order (see also the main text). This can be done in various efficient ways, and has the advantage that the gauge potential depends only on the system through the given $\Delta_{min}$ and $\Delta_{max}$, which might outperform the variational gauge potential if the action for the adiabatic gauge potential is dominated by excited states, resulting in a potential that is not expected to perform well for CD driving w.r.t. the ground state. 

\section{Variational minimization}
The exact gauge potential can be found by minimizing the action \cite{sels_minimizing_2017}
\begin{equation}
S(\chi) = \Tr \left[G^{\dagger}(\chi) G(\chi)\right], \  G(\chi) = \partial_{\lambda}\mathcal{H}-i[\mathcal{H},\chi],
\end{equation}
which is minimal precisely when $\chi = \mathcal{A}_{\lambda}$. This allows for the construction of approximate local gauge potentials by minimizing $S(\chi)$ in a restricted basis for $\chi$, including e.g. all local operators with a given support.

\begin{figure}
    \centering
    \includegraphics{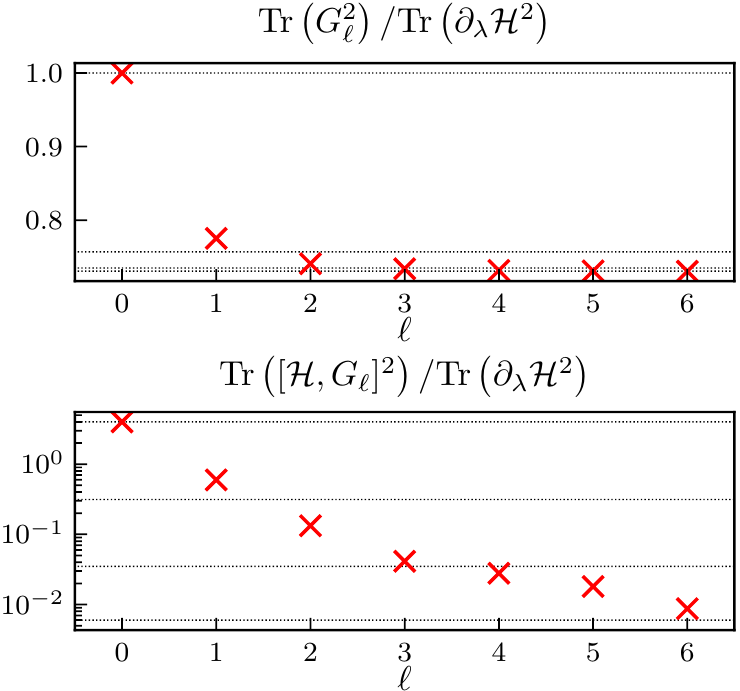}
    \caption{Variational minimum for $S_{\ell} = \Tr \left[G_{\ell}^2\right]$ and resulting error in $[\mathcal{H},G_{\ell}]$ for Fig.~1 in the main text. Everything is normalized by $\Tr\left[\partial_{\lambda}\mathcal{H}^2\right]$ in order to be system-size independent. The dotted lines denote the results when constructing the gauge potential in a local basis with support $d=1,2,3,4$.}
    \label{fig:var_GH}
\end{figure}

The results for the Figure in the main text are given in Fig.~\ref{fig:var_GH}, as compared with the variational results when constructing the gauge potential in a local basis as
\begin{equation}
\mathcal{A}_{\lambda} \approx \sum_{i=1}^L \alpha^{a_{1} a_{2} \dots a_{d}}_{i, i+1, \dots, i+d}\  \sigma_{i}^{a_1} \sigma_{i+1}^{a_{i+1}} \dots \sigma_{i+d}^{a_{i+d} },
\end{equation}
with $\alpha^{a_1 a_2 \dots a_d}_{i, i+1, \dots, i+d}$ the variational parameters and $a_i = x,y,z,0$. While the variational procedure minimizes $S_{\ell} = \Tr \left[G_{\ell}^2\right]$, the variational minimum is obtained when $[\mathcal{H},G_{\ell}]=0$, such that the latter can also be used as a measure for the resulting error. Despite only having a fraction of the parameters in the local ansatz, it is clear that the nested commutator ansatz can capture most of the relevant local contributions to the gauge potential.

An additional interpretation can be given to the minimization of the coefficients in $\mathcal{A}_{\lambda}^{(\ell)}$. Taking $\chi = \mathcal{A}_{\lambda}^{(\ell)}$, we can write $G_{\ell}\equiv G(\mathcal{A}_{\lambda}^{(\ell)})$ as
\begin{equation}
G_{\ell} = \partial_{\lambda}\mathcal{H}+\sum_{k=1}^{\ell} \alpha_k \underbrace{[\mathcal{H},[\mathcal{H},\dots [\mathcal{H}}_{2k}, \partial_{\lambda}\mathcal{H}]],
\end{equation}
where the action can again be expanded in the eigenbasis of $\mathcal{H}$ as 
\begin{equation}
S(\chi) = \sum_{m,n} |\braket{m|\partial_{\lambda}\mathcal{H}|n}|^2 \left[1+\sum_{k=1}^{\ell}\alpha_k \omega_{mn}^{2k}\right]^2.
\end{equation}
The resulting minimization leads to a linear set of equations
\begin{align}
&\sum_{k=1}^{\ell} \alpha_{k} \sum_{m,n} \omega_{mn}^{2(k+l)} |\braket{m|\partial_{\lambda}\mathcal{H}|n}|^2 \nonumber \\
&\qquad \quad = \alpha_{l} \sum_{m,n}\omega_{mn}^{2l} |\braket{m|\partial_{\lambda}\mathcal{H}|n}|^2, \ \  l=1 \dots \ell.
\end{align}
Defining the response function
\begin{equation}
\Gamma_{\lambda}(\omega)\equiv \Gamma(\omega,\partial_{\lambda}\mathcal{H}) = \sum_{m,n} \braket{m|\partial_{\lambda}\mathcal{H}|n}^2 \delta(\omega-\omega_{mn}),
\end{equation}
its moments follow as
\begin{equation}
\Gamma_{\lambda}^{(k)} = \int \mathrm{d}\omega\ \Gamma_{\lambda}(\omega) \omega^{2k},
\end{equation}
such that the linear set of equations can be rewritten as
\begin{equation}
\sum_{k=1}^{\ell} \alpha_k \Gamma_{\lambda}^{(k+l)} =  \Gamma_{\lambda}^{(l)}, \qquad l=1 \dots \ell.
\end{equation}
The left-hand side can be seen as the $(2l+1)$-th moment of $\sum_{k=1}^{\ell} \alpha_k \omega^{2k-1} \Gamma_{\lambda}(\omega)$, the response function of the approximate gauge potential, where the right-hand side can be seen as the $(2l+1)$-th moment of $\Gamma_{\lambda}(\omega)/\omega$, the response function for the exact gauge potential, such that the approximate gauge potential reproduces the first $\ell$ moments of the response function
\begin{equation}
\Gamma(\omega,\mathcal{A}_{\lambda})^{(k)} = \Gamma(\omega,\mathcal{A}^{\ell}_{\lambda})^{(k)}, k=1 \dots \ell.
\end{equation}

\section{Floquet Hamiltonian}
Here, we will calculate the matrix elements of the Floquet Hamiltonian defined in the main text. Consider the infinite-frequency limit of
\begin{align}
\mathcal{H}_{FE}(t) = \left[1+\frac{\omega}{\omega_0} \cos(\omega t) \right]\mathcal{H}(\lambda) +\dot{\lambda}\beta(t) \partial_{\lambda}\mathcal{H}(\lambda),
\end{align}
with
\begin{equation}
\beta(t) = \sum_{k=1}^{\infty}\beta_k \sin((2k-1)\omega  t).
\end{equation}
The leading-order contribution to the Floquet Hamiltonian can be obtained by first going to the rotating frame w.r.t. $\frac{\omega}{\omega_0} \cos(\omega t) \mathcal{H}(\lambda)$ as
\begin{align}
\tilde{\mathcal{H}}_{FE}(t)& = e^{i \frac{\sin(\omega t)}{\omega_0}\mathcal{H}(\lambda)} \mathcal{H}_{FE}(t) e^{-i \frac{\sin(\omega t)}{\omega_0}\mathcal{H}(\lambda)} 
\end{align}
from
\begin{equation}
\frac{\sin(\omega t)}{\omega_0}\mathcal{H}(\lambda) = \frac{\omega}{\omega_0} \int_{0}^t\mathrm{d}s \cos(\omega s)\mathcal{H}(\lambda),
\end{equation}
where we have assumed that $\mathcal{H}(\lambda)$ can be taken to be constant during a driving cycle. In this way, the rotating frame coincides with the lab frame at $t=0$ and $t=T$. The dominant contribution to the Magnus expansion is given by the time-averaged Hamiltonian in the moving frame as
\begin{align}
\tilde{\mathcal{H}}_{F}^{(0)}&=  \frac{1}{T}\int_{0}^T \mathrm{d}t \ e^{i \frac{\sin(\omega t)}{\omega_0}\mathcal{H}(\lambda)} \mathcal{H}_{FE}(t) e^{-i \frac{\sin(\omega t)}{\omega_0}\mathcal{H}(\lambda)}.
\end{align}
In order to continue, it will prove to be convenient to express the matrix elements in the eigenbasis of $\mathcal{H}(\lambda)$, where the off-diagonal elements are given by 
\begin{align}
\braket{m|\tilde{\mathcal{H}}_{F}^{(0)}|n} &= \frac{\dot{\lambda}}{T} \int_{0}^T \mathrm{d}t \ e^{i \sin(\omega t)\frac{(\epsilon_m-\epsilon_n)}{\omega_0}} \beta(t)\braket{m|\partial_{\lambda}\mathcal{H}|n} \nonumber \\
&= \dot{\lambda} \sum_{k=-\infty}^{\infty} \mathcal{J}_k\left(\frac{\epsilon_m-\epsilon_n}{\omega_0}\right)\braket{m|\partial_{\lambda}\mathcal{H}|n} \nonumber \\
&\qquad \qquad \times \frac{1}{T} \int_{0}^T \mathrm{d}t \ e^{i \omega k t}  \beta(t),
\end{align}
where the Jacobi-Anger formula has been used in order to recast the exponential as a sum of Bessel functions of the first kind. The integral then returns the Fourier coefficients of $\beta(t)$, leading to
\begin{equation}
\braket{m|\tilde{\mathcal{H}}_F^{(0)}|n} = i\dot{\lambda}\sum_{k=1}^{\infty} \beta_k \mathcal{J}_{2k-1}\left(\frac{\epsilon_m-\epsilon_n}{\omega_0}\right)\braket{m|\partial_{\lambda}\mathcal{H}|n}.
\end{equation}
Since the rotating frame coincides with the lab frame at initial and final times, the resulting Floquet Hamiltonian satisfies 
\begin{equation}
\braket{m|{\mathcal{H}}_F^{(0)}|n} =i \dot{\lambda} \sum_{k=1}^{\infty}\beta_k \mathcal{J}_{2k-1}\left(\frac{\epsilon_m-\epsilon_n}{\omega_0}\right)\braket{m|\partial_{\lambda}\mathcal{H}|n}.
\end{equation}
The contribution to the diagonal elements is simply given by the time-averaged $\mathcal{H}(\lambda)$, which can be assumed constant within a single driving cycle, leading to the proposed expression in the main text. The Bessel functions can be Taylor expanded around zero, leading to
\begin{align}
\mathcal{H}_F =& \mathcal{H}+i\dot{\lambda}\sum_{k=1}^{\infty}\beta_k \sum_{m=0}^{\infty} \frac{(-1)^m (2\omega_0)^{-2k-2m+1}}{m!(m+2k-1)!} \nonumber\\
&\qquad \qquad \qquad \times \underbrace{[\mathcal{H},[\mathcal{H},\dots [\mathcal{H}}_{2m+2k-1}, \partial_{\lambda}\mathcal{H}]].
\end{align}

The first-order correction on this Hamiltonian can also be calculated in the moving frame as
\begin{equation}
\mathcal{H}_F^{(1)} = \frac{1}{2iT^2}\int_{0}^T \mathrm{d}t_1 \int_{0}^{t_1}\mathrm{d}t_2\  [\tilde{\mathcal{H}}_{FE}(t_1),\tilde{\mathcal{H}}_{FE}(t_2)],
\end{equation}
which can be expanded as
\begin{align}
&\braket{m|\mathcal{H}_F^{(1)}|n}  = \frac{\dot{\lambda}}{2iT^2} (\epsilon_m-\epsilon_n) \braket{m|\partial_{\lambda}  \mathcal{H}|n} \nonumber \\
&\times\int_{0}^T \mathrm{d}t_1 \int_{0}^{t_1}\mathrm{d}t_2 \bigg[(1+\frac{\omega}{\omega_0}\cos(\omega t_1))\beta(t_2)e^{i\frac{\epsilon_m-\epsilon_n}{\omega_0}\sin(\omega t_2)} \nonumber \\
&\qquad \qquad \qquad\qquad\qquad\qquad\qquad- (1 \leftrightarrow 2) \bigg]\nonumber \\
&+\frac{\dot{\lambda}^2}{2iT^2} \sum_l \int_{0}^T \mathrm{d}t_1 \int_{0}^{t_1}\mathrm{d}t_2 \beta(t_1)\beta(t_2)e^{i\sin(\omega t_1)\frac{\epsilon_m-\epsilon_l}{\omega_0}} \nonumber \\
&\qquad \qquad \qquad \times e^{i\sin(\omega t_2)\frac{\epsilon_l-\epsilon_n}{\omega_0}} \braket{m|\partial_{\lambda}\mathcal{H}|l}\braket{l|\partial_{\lambda}\mathcal{H}|n}.
\end{align}
This can no longer be exactly evaluated because of the sum over the full Hilbert space, but it should be clear that the $\mathcal{O}(T)$ correction has two contributions determined by $\dot{\lambda}\braket{m|[\mathcal{H},\partial_{\lambda}\mathcal{H}]|n}$ and $\dot{\lambda}^2 \braket{m|\partial_{\lambda}\mathcal{H}|l}\braket{l|\partial_{\lambda}\mathcal{H}|n}$. This first term results in a correction on the coefficients in the dominant contribution, whereas the second term introduces new interactions in the Floquet Hamiltonian scaling as $\dot{\lambda}^2$.

\section{Examples}

In this Appendix, we explicitly calculate the single-commutator expansion for the two-qubit systems in the main text. First consider the two-level system
\begin{align}
&\mathcal{H} =  J \left(\sigma_1^x\sigma_2^x+\sigma_1^z\sigma_2^z\right) + h_z (\lambda-1)\left(\sigma_1^z+\sigma_2^z\right),\\
&\partial_{\lambda}\mathcal{H} =h_z \left(\sigma_1^z+\sigma_2^z\right).
\end{align}
The first-order commutator is given by
\begin{equation}
[\mathcal{H},\partial_{\lambda}\mathcal{H}] = -2 i  J h_z \left(\sigma_1^y\sigma_2^x+\sigma_1^x\sigma_2^y\right),
\end{equation}
and keeping only this term in the commutator expansion leads to
\begin{equation}
\mathcal{A}_{\lambda}^{(1)} = 2 \alpha_1 J h_z  \left(\sigma_1^y\sigma_2^x+\sigma_1^x\sigma_2^y\right).
\end{equation}
The single coefficient $\alpha_1$ follows from the operator $G_1 = \partial_{\lambda}\mathcal{H}-i[\mathcal{H},\mathcal{A}_{\lambda}^{(1)}]$, given by
\begin{align}
G_1 =&\ h_z \left(1+ \alpha_1 4 J^2 \right)\left(\sigma_1^z+\sigma_2^z\right) \nonumber \\
& - \alpha_1(\lambda-1)8  J h_z^2  \left(\sigma_1^x \sigma_2^x - \sigma_1^y \sigma_2^y\right),
\end{align}
leading to the action $S_1 = \Tr\left[G_1^2\right]$ as
\begin{align}
S_1 = 2 h_z^2\left(1+ \alpha_1 4 J^2 \right)^2+2\alpha_1^2 (\lambda-1)^2\left( 8 J h_z^2 \right)^2.
\end{align}
Minimizing $S_1$ leads to a linear equation for $\alpha_1$ and
\begin{equation}
\alpha_1 = -\frac{1}{4J^2+16(\lambda-1)^2h_z^2},
\end{equation}
resulting in the proposed gauge potential
\begin{equation}
\mathcal{A}_{\lambda}^{(1)} = -\frac{ J h_z}{2}\frac{\left(\sigma_1^y\sigma_2^x+\sigma_1^x\sigma_2^y\right)}{J^2+4(\lambda-1)^2h_z^2}  .
\end{equation}
For the three-level system, $\mathcal{A}_{\lambda}^{(1)}=i \alpha_1 [\mathcal{H},\partial_{\lambda}\mathcal{H}]$ can also be exactly calculated. Starting from
\begin{align}
&\mathcal{H} = -2J \sigma_1^z \sigma_2^z - h\left(\sigma_1^z+\sigma_2^z\right)+2 h \lambda \left(\sigma_1^x+\sigma_2^x\right),\\
&\partial_{\lambda}\mathcal{H} = 2 h \left(\sigma_1^x+\sigma_2^x\right),
\end{align}
the relevant operators follow as
\begin{align}
&[\mathcal{H},\partial_{\lambda}\mathcal{H}] = -8iJh \left(\sigma_1^y\sigma_2^z+\sigma_1^z\sigma_2^y\right) -4ih^2(\sigma_1^y+\sigma_2^y), \nonumber\\
&G_1= \left(2h+\alpha_1(32J^2 h + 8 h^3)\right)\left(\sigma_1^x+\sigma_2^x \right) \nonumber\\
&+ 32 \alpha_1 J h^2 \left(\sigma_1^x\sigma_2^z+\sigma_1^z\sigma_2^x\right) + 16 \alpha_1\lambda h^3 \left(\sigma_1^z+\sigma_2^z \right)\nonumber \\
&+ 64 \alpha_1J \lambda h^2 \left(\sigma_1^z\sigma_2^z-\sigma_1^y\sigma_2^y\right).
\end{align}
Minimizing the resulting action then returns
\begin{equation}
\alpha_1 = -\frac{J^2+h^2/4}{(4J^2+h^2)^2+(2\lambda h^2)^2+(4Jh)^2+(8J \lambda h)^2}.
\end{equation}

\bibliographystyle{apsrev4-1}
\bibliography{MyLibrary.bib}